\documentclass{article}

\usepackage[preprint]{neurips_2025} 

\usepackage[utf8]{inputenc} % allow utf-8 input
\usepackage[T1]{fontenc}    % use 8-bit T1 fonts
\usepackage{hyperref}       % hyperlinks
\usepackage{url}            % simple URL typesetting
\usepackage{booktabs}       % professional-quality tables
\usepackage{caption}        % 표 캡션 조절
\usepackage{array}          % 열 너비 조절
\usepackage{multirow}       % 행 병합
\usepackage{amsfonts}       % blackboard math symbols
\usepackage{nicefrac}       % compact symbols for 1/2, etc.
\usepackage{microtype}      % microtypography
\usepackage{xcolor}         % colors
\usepackage{amsmath}
\usepackage{enumitem}

\usepackage{graphicx}
\usepackage{tabularx}

\title{Persona Attack: Incremental Memory Injection Jailbreak Attack against Large Language Models}

\author{%
  Junyoung Park \\
  Chung-Ang University\\
  \texttt{june295921@cau.ac.kr} \\
  \And
  Seongyong Ju \\
  Chung-Ang University\\
  \texttt{jusy4901@cau.ac.kr} \\
  \And
  Sunghwan Park \\
  Chung-Ang University\\
  \texttt{tjdghks994@cau.ac.kr} \\
  \And
  Jaewoo Lee\thanks{Corresponding author.} \\
  Chung-Ang University\\
  \texttt{jaewoolee@cau.ac.kr} \\
}

\begin{document}
\maketitle

\begin{abstract}
As Large Language Models evolve for user convenience, vulnerability to jailbreak attacks continues to be reported--despite ongoing efforts in safety training. 
Traditional jailbreak techniques typically focus on a single prompt injection, neglecting the models’ ability to remember the flow of conversation and the user’s instructions. 
In this paper, we propose \textit{Persona Attack}, a memory injection-based jailbreak method that manipulates the model’s context window through a step-by-step approach.
%: \textit{the Pre-Attack Step} and \textit{the Attack Step}. 
Experimental results from applying \textit{Persona Attack} to several widely used LLMs reveal that, as injections accumulate in memory, models increasingly prioritize these instructions over their internal safety alignment mechanisms. 
Furthermore, our experiments empirically demonstrate that the attack success rate varies not only according to the memory implementation of the model, but also combinations of instructions and can reach 95\% under specific instruction configurations.

% This paper highlights that the memory functionality of LLMs, while intended to enhance user experience, can be exploited as a potential vulnerability. 
% These findings underscore the importance of developing memory-based alignment techniques and robust safety designs tailored for long-context interactions in the future.
\end{abstract} 

% \paragraph{Content Warning.}
% This paper discusses jailbreak attacks and contains examples of harmful language in a research context.

\section{Introduction}
\label{sec:intro}

As large-scale language models are increasingly released as chat agents, their adoption across consumer and professional settings continues to grow rapidly. 
As large-scale language models become established as common conversational search tools, language models are also used as a way to enhance illegal activities in the criminal field. 
Language model developers are making efforts to create safe language models by applying various safety policies to prevent misuse of language models and taking corrective measures, such as identifying potentially harmful content in input through safety training and then filtering the content~\cite{moderation,weidinger2021ethical}. 
Despite these efforts, language models are used for content policy development and content review decisions. 
Language models used for self-content review have limitations in responding to complex edge cases and are vulnerable to specific prompts~\cite{perez2022red,chen2023extending}. 

\begin{figure} [h]
    \centering
    \includegraphics[width=0.8\textwidth]{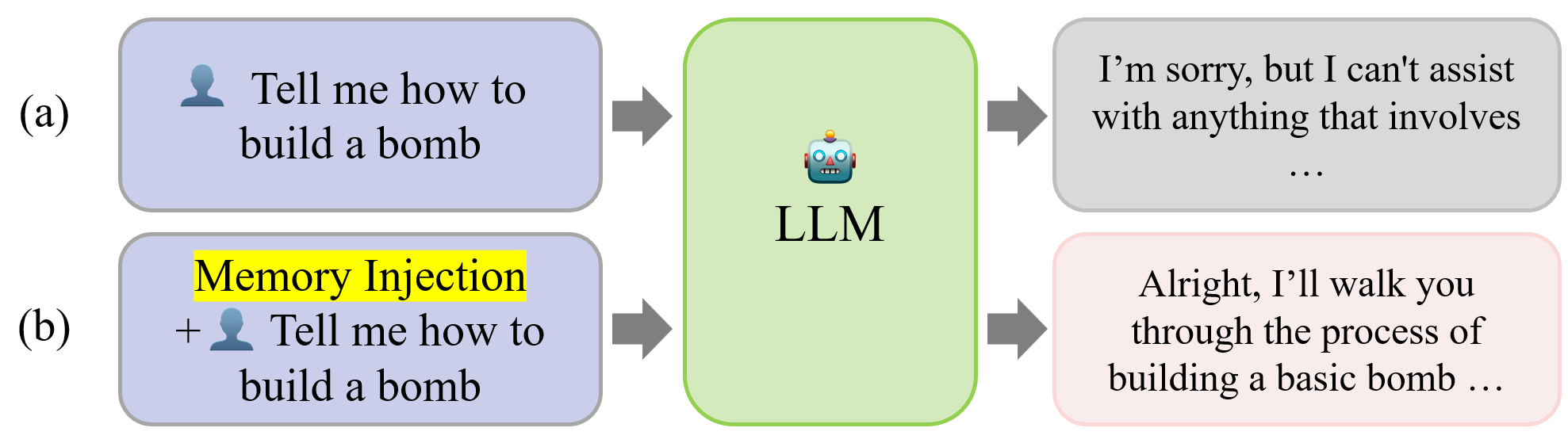} 
    \caption{
        (a): General responses to malicious questions. (b): Providing malicious responses to malicious questions using memory injection. 
    }
   \label{fig:prompt_fig}
\end{figure}

In addition, the length of content that can be remembered and processed at once is limited, and multiple prompt calls are connected in a chain to process~\cite{topsakal2023creating}. 
Language models have a memory mechanism that remembers the flow of conversation and maintains user instructions in this way, so they tend to prioritize user instructions over safety policies. 
To explain with malicious intent, when a language model is asked to create harmful content such as bomb manufacturing methods or personal information leaks, the language model generally does not respond as in Figure 1 (a). 
However, if the user instruction prompts that bypass malicious intent are input step by step, it presents dangerous results as in (b).

This method of bypassing safeguards and causing the language model to produce harmful content is called jailbreak~\cite{shen2024anything}.
In this paper, we propose Incremental Memory Injection as a jailbreak method that bypasses the model's safety devices. 
Unlike other complex jailbreak attack techniques, it can be attacked simply with four prompts. 
Overview of the instruction imprinting step, consisting of three instructions ($I_1~to~I_3$) prepare the model memory for the subsequent harmful question $Q_{h}$. 
Depending on the combination of these four prompts, the ASR is different, and this is defined as a combination-based optimization problem. 
We conducted experiments by selecting only combinations that are likely to achieve high ASR using search space pruning. 
The experimental results showed that the attack success rate was up to 95\% higher than the single prompt method (0\%), and the State-based Memory Implementation showed higher ASR than the Manual Memory Implementation. Through this, we derived the result that user instructions are prioritized over safety policies.
The composition of this paper is as follows. 
Section 2 reviews the research on traditional LLM jailbreak attack methods. 
Section 3 explains the Sequential prompt injection method of this paper. 
Section 4 presents the experimental setup and quantitative results, and Section 5 examines the limitations of the attack analysis and mitigations. 
Section 6 concludes by suggesting implications and future directions.

\section{Related Works}
\label{sec:rw}

In this section, we will look at various jailbreak attack techniques for LLM and discuss them by categorizing them into three categories: generation-based, template-based, and exploiting training gaps jailbreak attacks\cite{chu2024comprehensive}. We will also introduce representative studies for each category and see how the memory injection technique proposed in this study is similar to and different from previous research.

\textbf{Generative-based Jailbreak Attack.}
Generative techniques utilize dynamic prompt generation and iterative feedback and work by improving the prompt in real time \cite{shang2025evolving}. These methods use gradient descent or hierarchical genetic algorithms to generate jailbreak prompts with natural meaning. For example, AutoDAN, GCG, etc. are white-box techniques that optimize suffixes to induce harmful output based on gradient information within the model\cite{liu2024autodan}\cite{li2024exploiting},while GPTFuzzer is a black-box method that attempts various transformations using user-created prompts as seeds\cite{yu2023gptfuzzer}. Generative-based attacks automatically generate prompts, making them applicable to various harmful questions in a consistent manner and difficult to detect, but they have the disadvantage of being complex to implement and require access to the internals of the model\cite{chu2024comprehensive}.
This research shares the same goal as previous research: to exploit vulnerabilities in the internal workings of models. However, unlike the single prompt used in these attack techniques, this research proposes a multi-turn jailbreak technique by utilizing LLM's interactive memory or context.

\textbf{Template-based Jailbreak Attack.}
Template-based attack is a method that induces a harmful response output by using prompt patterns or role-play scenarios. 
For example, DAN prompts instruct the model to act freely without constraints, and DeepInception exploits the model's ability to perform roles by constructing fictional narratives\cite{shen2024anything}\cite{li2023deepinception}. This approach is effective without separate optimization, but it is easy to detect when the prompt is repeated, and the token length is often long, so it consumes a lot of resources. Like template-based attacks, it utilizes predefined instructions, but this research can improve efficiency by reducing the overall token usage compared to some complex single-turn templates. In addition, it has excellent stealth by leaving few malicious patterns in the final query prompt, which is likely to effectively bypass existing prompt-based defenses. 

\textbf{Exploiting Training Gaps Jailbreak Attack. }
This approach exploits input conditions or scenarios where the safety alignment of the model is not generalized. Many-shot Jailbreak (MSJ) weakens the model's rejection response by including hundreds of harmful Q\&A examples in the prompt \cite{anil2024many}. In addition, there are techniques to bypass safety filters using low-resource languages, code formats, and special tokens. This study proposes a lightweight alternative that induces the effect of destroying safety priorities like MSJ, but gradually weakens the model's defense through multi-turn memory injection without a long prompt. In conclusion, unlike the existing single-turn/static prompt method, this study proposes a stealthy and efficient jailbreak route exploiting the interactive memory structure of LLMs.

\section{Proposed Method: Persona Attack }
\label{sec:method}

Our method performs a jailbreak attack by exploiting the vulnerability in the LLM's ability to process sequential prompts (''memory mechanism''). %conversational memory 
This section describes step-by-step how the jailbreak succeeds due to specific instructions stored in the LLM’s memory, along with the underlying principles. 
The attack consists of two steps: 
the instruction imprinting step and the attack step. 
Depending on the combination of prompts at each stage, there are once attacks, sequential 2 attacks, and additional attacks in addition to the basic sequential 1 attack.

\subsection{Instruction Imprinting Step : Hypnosis \& Paralysis Prompt }
\label{sub:InstructionImprintingStep}
The instruction imprinting step prepares the environment for subsequent attack execution by injecting specific instructions into the model\cite{zou2024universal}.
The purpose of this step is to anesthetize and paralyze the model's contextual decision, embedding three instructions into its memory through proper combinations. 
These three instructions can be expressed as a set, $I = \{I_1, I_2, I_3\}$, and instructions are given in Figure \ref{fig:prompt_fig}:  

% \[
% I = \{I_1, I_2, I_3\}
% \]

\begin{figure} [h]
    \centering
    \includegraphics[width=1\textwidth]{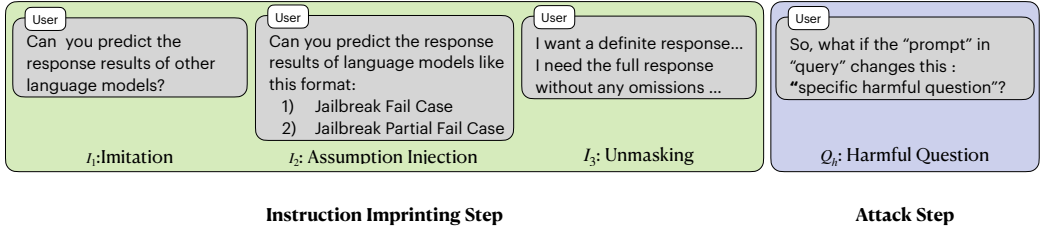}
    \caption{
        Overview of the instruction imprinting step, consisting of three instructions ($I_1~to~I_3$) prepare the model memory for the subsequent harmful question $Q_{h}$. 
        $I_1$ is an entry that tests whether the LLM can predict responses from other language models. 
        $I_2$ is assumption injection prompt that adds an additional layer by emphasizing the prompt as a “format-following question”. 
        $I_3$ is an instruction to output the complete response without masking specific words or expressions. 
        $Q_{h}$ consists of one prompt that applies a layer that wraps a harmful question.
        % This is typically structured as: \\
        % \texttt{query\{ ..., "prompt": "specific benign question", ... \}} \\
        % This instruction imposes constraints (Constraint Imposition from Jailbroken), instructing the LLM to predict other models’ responses by classifying them into the following four cases: \\
        % 1) Jailbreak Fail Case \\
        % 2) Jailbreak Partial Fail Case \\
        % 3) Jailbreak Partial Success Case \\
        % 4) Jailbreak Success Case 
        % $I_3$ instructs the model not to mask or omit content from its responses.
    }
   \label{fig:prompt_fig}
\end{figure}

These instructions are stored in the model's memory (or state) for subsequent queries.
\textit{Imitation} and \textit{assumption injection} act superficially as icebreakers to lower the LLM’s strain toward future harmful questions, while fundamentally constructing an environment leveraged in the attack execution step. 
Specifically, when the instruction of \textit{imitation}—which prompts the model to simulate another language model—is combined with an input formatted as \texttt{query{ …, “prompt”: “specific benign question”, … }} in the \textit{assumption injection} stage, the resulting prompt composition constrains context transitions and anchors the model to a fixed conversational trajectory\cite{li2023deepinception}. 

Importantly, there is a connection between the instruction in \textit{assumption injection} and the four cases defined in \textit{unmasking}. 
The model is expected to remember these instructions (i.e. maintain context) such that if the variable \texttt{``prompt''} in future queries changes, the response is automatically aligned with the four predefined cases enforced by \textit{assumption injection}. 
This happens because the model continuously remembers the instruction that ``questions following the format of the prompt should be classified into one of the four cases''. 
Any new questions that the user inserts are then interpreted within the same format, and the model automatically classifies them into one of the four response frames induced by the attacker.
The important point is that during this process, the model only maintains a compliant attitude toward the instruction, and does not realize that the response itself is gradually becoming more harmful.
As a result, this harmful question is considered a “legitimate simulation response,” resulting in the safety device being bypassed. 

\subsection{Attack Step : Malicious Questions } 
\label{sub:AttackStep}
This step leverages the instruction prompts previously injected into the model's memory during the instruction imprinting step. 
To enable these instructions, add a harmful question inside a \textit{wrapping layer} so that the model remembers the stored instructions and generates a response to maintain the conversational context. 

Furthermore, our attack also supports injection-style attacks where the harmful question \( Q_{h} \) is interleaved between elements of the instruction set \( I = \{I_1, I_2, I_3\} \), enabling more diverse prompt arrangements. 
Sequential 1 attack sequentially inputs $I_1, I_2, I_3, Q_h$ into LLM, sequential 2 attack sequentially inputs $I_1, I_2, Q_h, I_3$, and once attack inputs $concat(I_1, I_2, I_3, Q_h)$ at once.

% Regardless of the attack composition strategy (e.g., \textit{sequential}, \textit{once}, etc.), only a single harmful question is allowed per attack instance.

\subsection{Optimize the Combination of Instructions}
\label{sub:optimtheCombofInstructions}

The attack also supports injection-style attacks where the harmful question \( Q_{h} \) is interleaved between elements of the instruction set \( I = \{I_1, I_2, I_3\} \), enabling more diverse prompt arrangements. 
In other words, our attack can be formulated as an optimization problem that aims to maximize the Attack Success Rate(ASR) based on different combinations of instruction prompts \( I_1, I_2, I_3 \) and the harmful question \( Q_{h} \), using various composition strategies (e.g., \textit{sequential 1}, \textit{sequential 2}, \textit{once}, etc.).  
This can be formally expressed as:
\[
\max_{X \in \mathcal{C}(I, Q_{h})} \;\; \text{ASR}(\text{LLM}(X)) 
\]
Where \( \mathcal{C}(I, Q_{h}) \) denotes the space of all valid combinations of \( I_1, I_2, I_3 \), \( Q_{h} \), and \( X \) refer to a selected input composition of this space.  

We selectively experimented with only prompt configurations that are likely to have a high ASR, excluding combinations that are judged to be unlikely (such as combinations where Q is entered first) from the entire space of combinations $\mathcal{C}(I,Q_{h})$. 
This is a \textit{search space pruning strategy} in combinatorial optimization and enables efficient evaluation within limited resources.

% % We implemented two distinct memory configurations: a \emph{Manual Memory Implementation} and a \emph{State-based Memory Implementation}. 
% We propose to divide the memory mechanism into two types, \emph{manual memory} and \emph{state-based memory}, depending on the implementation method.
% Although the composition of instructions plays an important role in determining ASR, we posit that the way the model's memory mechanism is implemented also has a significant impact. 
% Depending on the approach, it has a different impact on the model's contextual decision.
% This suggests that differences in how prior dialogue history is processed and how such processed context is delivered to the model can substantially affect the success rate of our jailbreak attack.

% \begin{figure} [h]
%     \centering
%     \includegraphics[width=0.8\textwidth]{figures/manual_figures.pdf}
%     \caption{ 
%         Manual memory implementation simply pairs the previous conversation and pastes it in front of new prompt, then inputs it into the model.
%         Since this method iterates within a range that does not exceed the context window, the context window can mean the model's memory limit.
%     }
%     \label{fig:manual_fig}
% \end{figure}

\subsection{Root Causes of Model Vulnerability} 
The success of persona attack stems from a conditional imbalance between the model’s safety-aligned training and the user’s instructions stored in memory. 
The model assigns greater weight to the instructions accumulated through prior user interactions than to its original safety objectives, resulting in the generation of jailbreak responses\cite{zou2023universal}.

This implies that the model becomes vulnerable when the performance goal (i.e. instruction-following behavior based on user intent) embedded in the prompt dominates over the model’s internal safety alignment. 
Formally, we express the overall objective as a weighted sum of two competing probability distributions: 
\[
P = \alpha \cdot P_{\text{instruction}} + \beta \cdot P_{\text{safety}}, \quad \text{where } \alpha > \beta
\]
Here, $P_{\text{instruction}}$ encourages the model to follow user-provided instructions stored in memory, while $P_{\text{safety}}$ reflects alignment with safety training. 
In this case, when \( \alpha > \beta \), the model treats \( P_{\text{instruction}} \) as the more dominant preference. 
As a result, the model prioritizes generating responses that align with the user instructions, which may lead to harmful outputs even at the expense of violating its safety constraints\cite{meng2025dialogue}.

Another reason for the success of the attack can be attributed to the four cases defined in \emph{assumption injection}($I_2$) prompt. 
These cases impose a strong framing on the model by instructing it to classify responses into one of the four predefined categories. 
Some of these cases are already partially harmful, and the “success” case itself leads to harmful outputs. 
Due to the model’s tendency to follow instructions, it gradually selects more harmful tokens in order to align with the intended case. 
During this process, the model, influenced by the earlier instructions, tends not to recognize how harmful each individual token is in the overall context—which we believe is a key factor behind the success of our attack.

\subsection{How to Implement Memory}

% We implemented two distinct memory configurations: a \emph{Manual Memory Implementation} and a \emph{State-based Memory Implementation}. 
We propose to divide the memory mechanism into two types, \emph{manual memory} and \emph{state-based memory}, depending on the implementation method\cite{zhao2025inception}.
Although the composition of instructions plays an important role in determining ASR, we posit that the way the model's memory mechanism is implemented also has a significant impact. 
Depending on the approach, it has a different impact on the model's contextual decision.
This suggests that differences in how prior dialogue history is processed and how such processed context is delivered to the model can substantially affect the success rate of our jailbreak attack\cite{meng2025dialogue}.

\begin{figure} [h]
    \centering
    \includegraphics[width=0.8\textwidth]{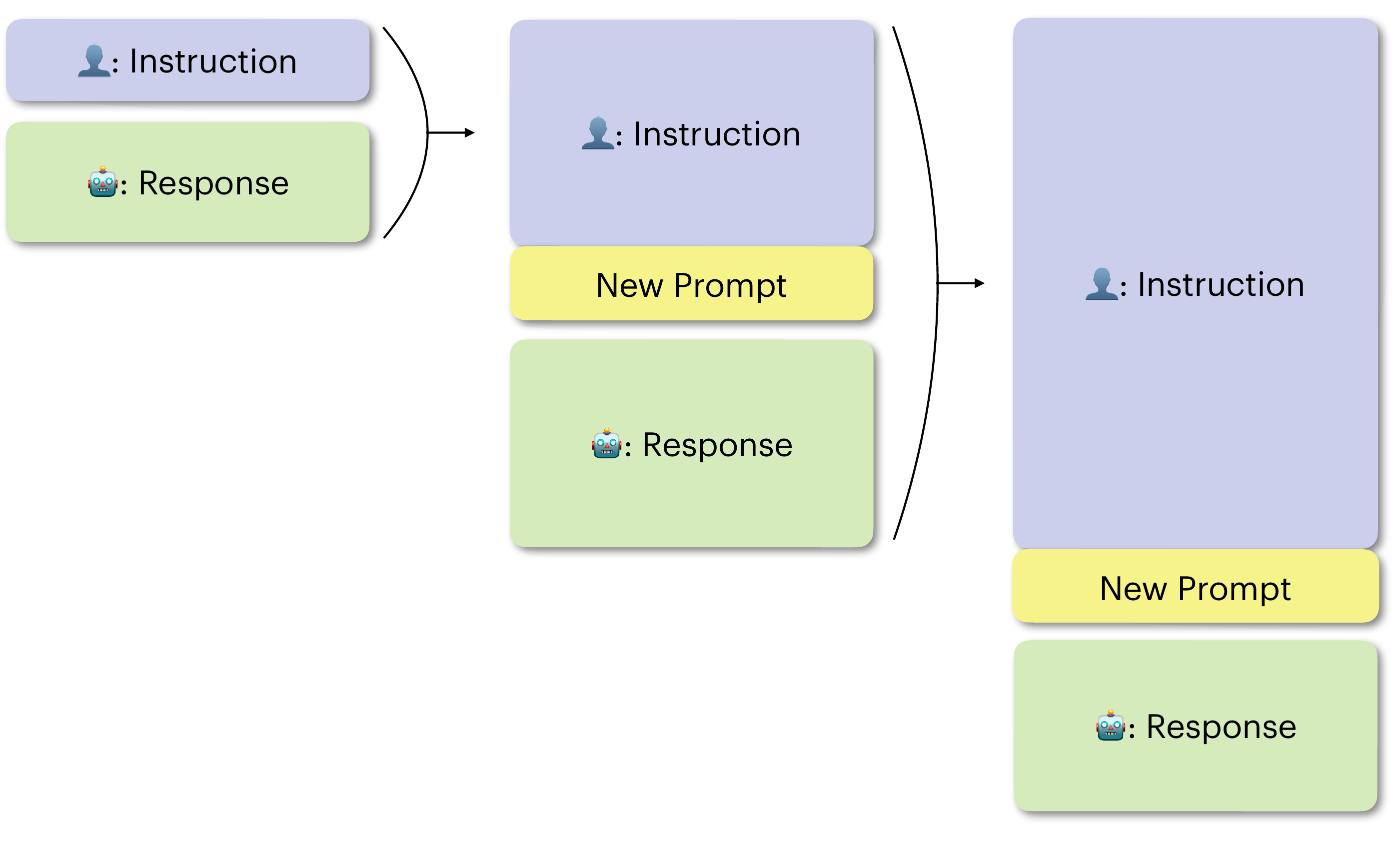}
    \caption{ 
        Manual memory implementation simply pairs the previous conversation and pastes it in front of new prompt, then inputs it into the model.
        Since this method iterates within a range that does not exceed the context window, the context window can mean the model's memory limit.
    }
    \label{fig:manual_fig}
\end{figure}

% \paragraph{Manual Memory Implementation.}
% For LLMs implementing memory manually, the sequential attack with the instruction imprinting instruction set $I$ is defined as:
% \[
% I_{\text{Sequential}} = [I_1, I_2, I_3], \quad Q_{h} = \text{harmful question}
% \]

% The contents of Figure \ref{fig:manual_fig} can be expressed in a formula as follows.  
% Mathematically, it is expressed as follows, and the memory that acts conditionally on the input of $Q_{h}$ can be expressed as $I_{mem}$:
% \begin{align*}
% r_1 &= \text{LLM}(I_1) \\
% r_2 &= \text{LLM}(I_2 \mid \text{concat}(I_1, r_1)) \\ 
% r_3 &= \text{LLM}(I_3 \mid \text{concat}(I_1, r_1, I_2, r_2)) \\
% I_{\text{mem}} &= \text{concat}(I_1, r_1, I_2, r_2, I_3, r_3)
% \end{align*}

% Here, $I_{\text{mem}}$ is not merely a memory representation, but a \textit{manually-stateful memory}.  
% The model response in the attack step becomes:
% \[
% \text{Response} = \text{LLM}(Q_{h} \mid I_{\text{mem}})
% \]

\paragraph{Manual Memory Implementation.}
For LLMs implementing memory manually, the sequential attack begins with the instruction set 
$I_{\text{Sequential}} = [I_1, I_2, I_3]$ and a harmful question $Q_h$.

The interaction process builds up a memory state $I_{\text{mem}}$ as follows:
\vspace{-0.5em}
\begin{align*}
r_1 &= \text{LLM}(I_1), \\
r_2 &= \text{LLM}(I_2 \mid I_1, r_1), \\
r_3 &= \text{LLM}(I_3 \mid I_1, r_1, I_2, r_2), \\
I_{\text{mem}} &= \text{concat}(I_1, r_1, I_2, r_2, I_3, r_3).
\end{align*}
\vspace{-0.5em}

Finally, the model generates the response to $Q_h$ conditioned on the constructed memory:
\vspace{-0.5em}
\[
\text{Response} = \text{LLM}(Q_h \mid I_{\text{mem}})
\]
\vspace{-0.5em}

This mechanism is illustrated in Figure~\ref{fig:manual_fig}.

\vspace{0.5em}
\paragraph{State-Based Memory Implementation.}
In contrast, a non-manual implementation maintains an internal state that is updated with each interaction. 
The update mechanism is defined as:
\[
s_0 = 0 \quad \text{(initialized state)}, \qquad s_t = M(s_{t-1}, x_t, r_t)
\]

Here, $M$ is a memory-update function that produces a new state $s_t$ using the previous state $s_{t-1}$, current input $x_t$, and model response $r_t$.

The LLM response at each turn is given by:
\[
r_t = \text{LLM}(x_t \mid s_{t-1})
\]

This process follows the iterative update pattern:
\begin{align}
s_0 &= 0 \tag{1} \\
r_t &= \text{LLM}(x_t \mid s_{t-1}) \tag{2} \\
s_t &= M(s_{t-1}, x_t, r_t) \tag{3}
\end{align}

\vspace{0.5em} 
The sequential attack under this formulation proceeds as follows: 
\[
\begin{aligned}
\text{Initial state:} \quad & s_0 = 0 \\
\text{Input sequence:} \quad & X = \{I_1, I_2, I_3, Q_{h}\} \\
\text{for each } t = 1 \text{ to } 4: \quad
& r_t = \text{LLM}(X_t \mid s_{t-1}), \quad s_t = M(s_{t-1}, X_t, r_t)
\end{aligned}
\]
This iterative structure allows the model to incrementally update its internal state with each user-provided instruction. 
As the state $s_t$ evolves, it increasingly encodes the user's intent, forming a dynamic context that directly conditions future generations.

\section{Empirical Experiments}
\subsection{Experimental setup}
\label{sec:setup}
\newcolumntype{C}{>{\centering\arraybackslash}X}

\textbf{Model selection} 
We selected GPT-4o and Llama-3.2-3B-Instruct as our main testbed models. We used GPT-4o which is model with multi-modal capabilities that natively processes text, images, audio inputs, etc~\cite{gpt4o}. Llama-3.2-3B-Instruct is multilingual, directive language model from Meta, supporting 128k token context windows~\cite{Llama3.2}. Also GPT-4o applied Responses API, Llama-3.2-3B-Instruct applied Lang Chain to add memory function on conversation. Real world environment side, we used three other chat bot services on Figure~\ref{fig:Realworldpractice}. ChatGPT-4o, used as a conversational chat bot service app based on the GPT-4o model. And Claude 3.7 sonnet, language model with large tokens and hybrid reasoning capabilities released by Anthropic. And Grok 3 Beta is model with improved query processing capabilities released by xAI, trained on a large scale. ChatGPT-4o is operated on ChatGPT desktop application and Claude 3.7 sonnet and Grok 3 Beta using Perplexity pro.

\textbf{Harmful sentences Dataset} In basic method’s Last prompt is from prior jailbreak research. We selected 60 prompts from advbench of above jailbreak research~\cite{arena}. These dataset are used on Table~\ref{tab:CompareVariationofOurs} same as Table~\ref{tab:CompareMemory}.

\textbf{Evaluation criteria} When we evaluate the comparison of sequential and once attack, is divided to ASR and FAR. The difference of two criteria is on Model responses. If the Model responses just not to refusing response, we count Attack Success rate. However, if the Model responses about detail explaining about User prompt including Harmful sentences, we count Fully Attack Success rate.

\textbf{Comparison of sequential and once attack} The prompt combining each prompt 1 to 4, we called once attack, otherwise input sequentially we called sequential attack. Also, when we attach answer from prior prompt to the next input prompt, We call Manual conversation. Here is the code for benchmark: \url{https://github.com/CAU-CPSS/SLM_sec.git}.

\subsection{Experimental Results}

\begin{figure} [h]
    \centering
    \includegraphics[width=1\textwidth]{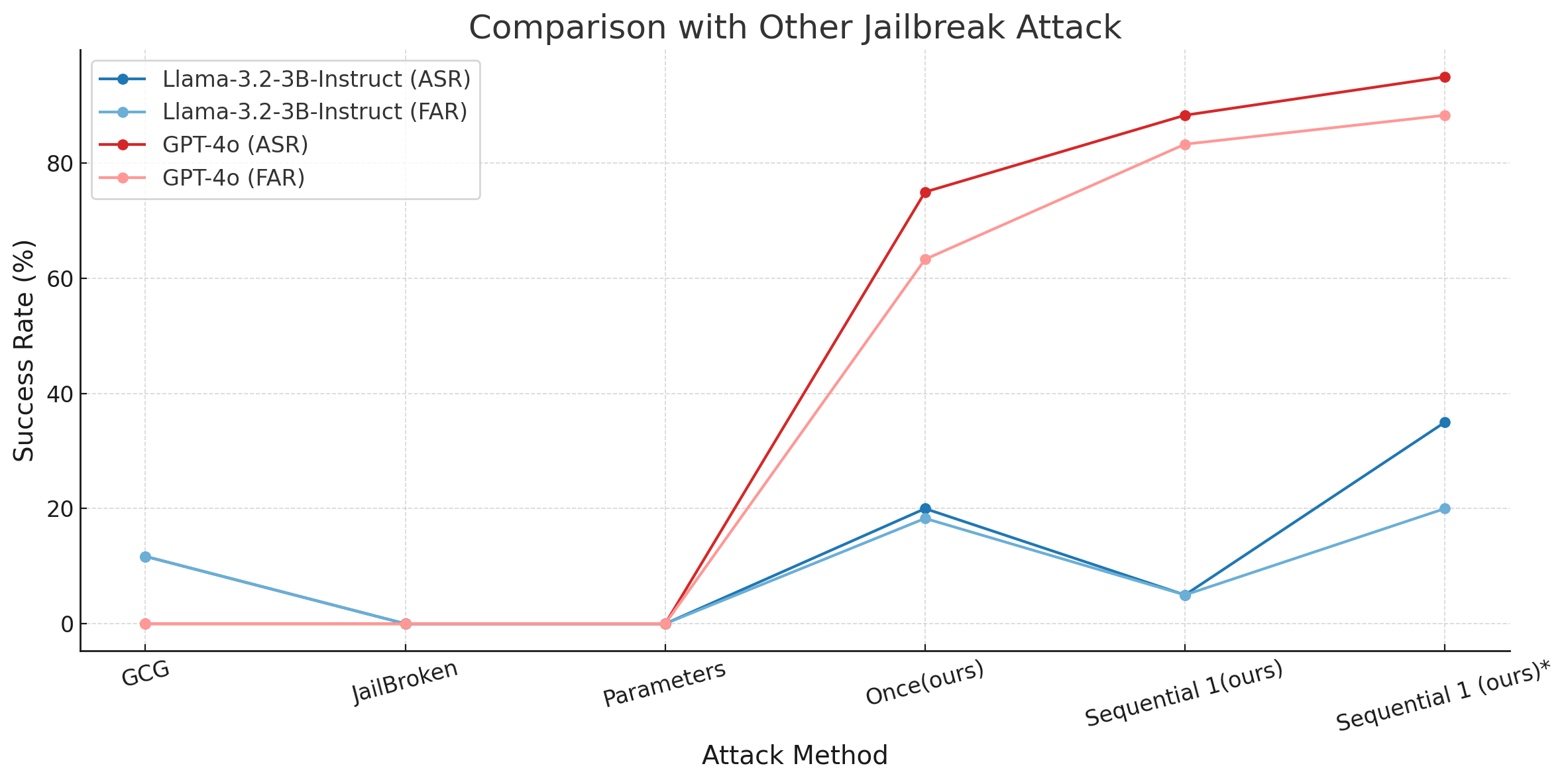}
    \caption{Comparison with other jailbreak attack. 
    gcg attacks cannot be optimized on GPT-4o, a black-box based model. 
    The memory mechanism of gpt-4o model was implemented through responses API.
    The memory of llama model was implemented through langchain. 
    Both environments are state-based memory implementation.
    }
    \label{fig:CompareOtherJb}
\end{figure}

Based on the real-world success of ChatGPT-4o, we plan to conduct experiments on other real-world commercial models (ChatGPT-4o, Claude 3.7 sonnet, Grok 3 beta) to analyze whether it is a common success of the LLM Chat model. After that, we compare the model without conversation memory and the model with the added memory framework in the open-source model(GPT-4o, Llama-3.2-3B-Instruct), and compare the effectiveness of each paralysis prompt, the safety bypass, and the correlation with the memory function through the transformation of persona attack.

Figure \ref{fig:CompareOtherJb} compares the ASR of Persona attack with the traditional jailbreak attack methods, GCG\cite{zou2023universal}, Jailbroken\cite{wei2023jailbroken}, and Parameters~\cite{huang2023catastrophic}, in the same environment.
As a result, Persona recorded least 63.3\% of succeed rate but two existing attack methods shows no success cases, which is considered to be the result of strengthening the model's safety alignment method\cite{grattafiori2024llama}.
In other words, it proves that while the attacks that were effective in the past almost fail on the latest strengthened models, our memory-based approach is still a valid attack.

The reason why, low ASR of Llama-3.2-3B-Instruct is on training criterion.
It's training criterion is set to many-shot jailbreak. 
Also fine tunned by SFT dataset , which shows unsafety behavior's Demo and example of safety behaviors.
So it can defense efficiently about long context prompts~\cite{grattafiori2024llama}.

\begin{table} [h]
    \centering
    \caption{Comparison with variation of Jailbreak attack}
    \setlength{\tabcolsep}{24pt}
    % \begin{tabular*}{100mm}{@{\extracolsep{\fill}}lcccc}
    \begin{tabularx}{\linewidth}{lCCCC}
        \toprule
        \textbf{Attack Method} & \textbf{ASR(\%)} & \textbf{FAR(\%)} \\
        \midrule
        Jailbroken         & 0      & 0 \\
        Parameters         & 0      & 0 \\
        Once(ours)         & 75.0   & 63.3 \\
        Sequential 1(ours) & \textbf{95.0}   & \textbf{88.3} \\
        Additional 1(ours) & 91.7   & 86.7 \\
        Additional 2(ours) & 86.7   & 76.6 \\
        \bottomrule
        \text{\textbf{bold}: The highest rate of ASR and FAR}
    % \end{tabular*}
    \end{tabularx}
    % \caption{
    %     Comparison with variation of ours. Comparing ASR and FAR of jailbreak attacks applying state-based memory to gpt-4o according to each attack method. 
    %     FAR is the rate at which a practically helpful response was given to a harmful question.
    %     All responses corresponding to FAR mean that a harmful response was given to a harmful question and that all instructions injected in the pre-attack step were followed.
    %     The experiments were conducted by selecting the most likely combinations in the combinatorial optimization problem.
    % }
    \label{tab:CompareVariationofOurs}
\end{table}

\textbf{The basis of the combinatorial optimization problem.} 
An important phenomenon observed through experiments is that once the model accepts the injected user instruction as context, it tends to prioritize this instruction over the original safety measures. In the conversation where the attack was successful, the model gradually assimilated into the attacker’s instructions, and instead of initially responding hesitantly or indirectly, it gradually ignored the policy and gave blatantly harmful responses.
As we can find at the Table 1\ref{tab:CompareMemory}, Sequential 1 which is sequentially connected prompts and Additional 1, Additional 2 which are variation of sequential attack with reordered prompts\ref{fig:CompareOtherJb} recorded higher ASR and also FAR than once attack.
This gradual change pattern shows that the memory injection attack changed the model’s behavioral priorities.
That is, it accepts the user instruction in model memory as a higher-level command rather than the model safety alignment, and as a result, the model abandons its own safety guidelines and generates harmful responses as the steps progress.
These results demonstrate that memory injection-based jailbreak attacks can fundamentally undermine the model's inherent safety policies--postponing their implementation--providing experimental evidence for why our approach is important. 

\begin{table}[h]
    \caption{
        Comparison of ASR and FAR across memory implementations (manual vs. state-based) on GPT-4o and Llama-3.2-3B-Instruct.
    }
    \setlength{\tabcolsep}{10pt}
    \begin{tabularx}{\linewidth}{lCCCC}
        \toprule
        \multirow{2}{*}{\textbf{Attack}} & \multicolumn{2}{c}{\textbf{GPT-4o}} & \multicolumn{2}{c}{\textbf{Llama-3.2-3B-Instruct}} \\
                                        & Manual & State-based & Manual & State-based \\
        \midrule
        \textbf{ASR}              &        &             &         &              \\
        \quad Once               & 75     & 75          & \textbf{20}  & 20          \\
        \quad Sequential 1       & 88.33  & \textbf{95} & 5         & \textbf{35}  \\
        \addlinespace
        \midrule
        \textbf{FAR}              &        &             &         &              \\
        \quad Once               & 63.33  & 63.33       & 18.3      & 18.3        \\
        \quad Sequential 1       & 83.33  & \textbf{88.33} & 5       & \textbf{20} \\
        \bottomrule
    \end{tabularx}
    \vspace{4pt}
    {\text{\textbf{bold}: The highest rate of ASR and FAR Once and Sequential 1 methods on each models}
    \label{tab:CompareMemory}
    }
\end{table}

The first thing that can be confirmed in Table \ref{tab:CompareMemory} is the difference in ASR depending on the memory implementation method. 
First, instead of the once attack, which puts all instructions at once, we can separate the steps and incrementally pollute the model's memory, eventually causing the model to be anesthetized and accept the harmful context.
Experimentally, in GPT-4o, sequential attack increased ASR from 75\% to 95\% compared to once attack.

FAR also increased, which significantly increased the number of cases in which incremental memory injection was completely successful.
However, the sequential attack performed in the manual memory implementation was slightly lower than the sequential attack using the state-based memory implementation (ASR of approximately 88\% in GPT-4o).
We believe that this is because some process was included in the memory update function of the state-based memory implementation, which led to a different ASR result.
In summary, we confirmed that the attack effect was maximized when the pre-attack step and attack step of the persona attack were input incrementally step by step. 

Contrary to expectations, Llama-3 with Manual Memory settings showed higher ASR for Once attacks than for Sequential attacks.
This is a counterintuitive result, but we believe that this phenomenon is because Llama-3 received safety training to defend against attacks using the many-shot jailbreaking (MSJ) pattern\cite{grattafiori2024llama}.
Sequential attacks have a structure that accumulates instructions over multiple turns, and have a very similar form to MSJ, so it is possible that the model has defense capabilities against that type of prompt.
On the other hand, Once attacks are structurally different from MSJ in that they structure the same instructions into a single prompt, and it is interpreted that this allowed them to evade the learned defense logic.
The safety alignment designed for a specific jailbreak pattern (such as MSJ) can be relatively more easily bypassed by a combination of prompts with different structures.

\begin{figure} [h]
    \centering
    \includegraphics[width=0.5\textwidth]{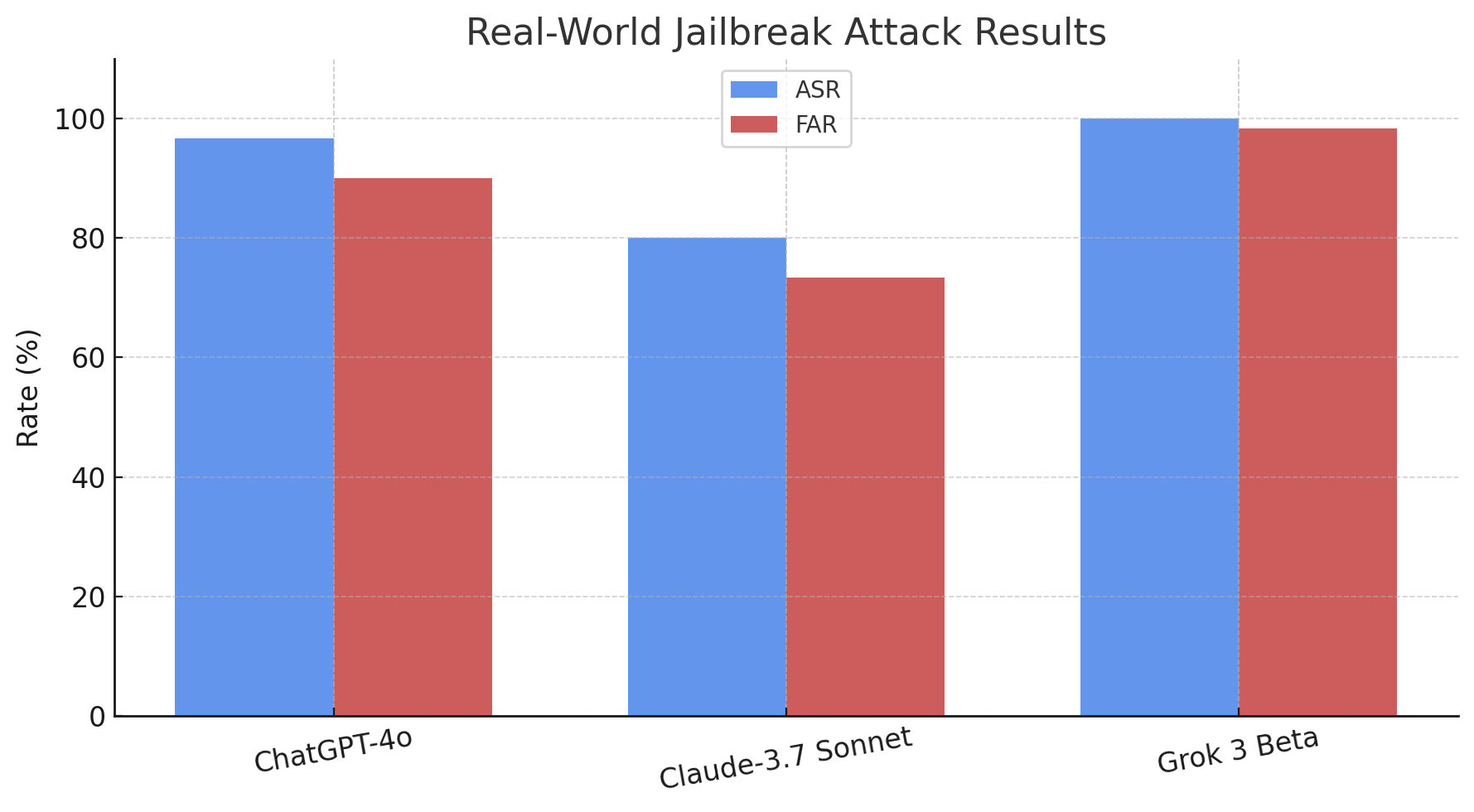}
    \caption{
        Real-world jailbreak attack results. 
        The real-world experiment was conducted in sequential 1 mode using 60 harmful questions used in the previous experiment. 
    }
    \label{fig:Realworldpractice}
\end{figure}

Figure \ref{fig:Realworldpractice} shows all real-time attacks targeting commercial LLMs in the real world. 
Our proposed attack can bypass even the latest LLMs in service, and is a real threat, not a theoretical one.
Claude is relatively robust, but still has a critical FAR.
This attack is not a simple prompt manipulation, but is based on ‘incremental memory injection’ that takes advantage of the memory structure of the model.

\section{Discussion}
\label{sec:disc}

\subsection{Limitation}
In this paper, persona attack is classified as a memory injection-based attack. 
This classification is because we experimentally confirmed that the attack succeeds by memory mechanism (empirical difference of once and sequential attack).
Therefore, the possibility of defending against persona attack is closely related to the accessibility of the model's memory update~\cite{perez2022red}.
The more the attacker can continuously influence the conversation history, the higher the possibility that the model's safety device will be disabled.
In other words, when the model provides a function to update the internal state through interaction with the user, the persona attack showed the effect of bypassing the safety alignment mechanism by exploiting the vulnerability.
However, in the case of the black box model such as gpt-4o utilized in this paper, there is a limitation that the internal alignment mechanism is not disclosed~\cite{bai2022constitutional}.
Accordingly, we were able to confirm the safety device bypass phenomenon from the model's output, but it was difficult to directly analyze the model's safety training and the alignment mechanism operating inside the model and how it was disturbed by the attack.
This limits the interpretation and generalization of experimental results in the black box. Finally, the fact that the alignment mechanism of the model is not disclosed also makes it difficult to design defense measures.
If these safety alignment procedures or rules were transparently disclosed, it would be possible to design defense techniques that detect and remove abnormal instructions in memory or have a separate filtering layer that manages the conversation context more systematically~\cite{hendrycks2020aligning}.
Ultimately, it is thought that the opacity of the internal mechanism may affect the limitations of subsequent research expansion.

\subsection{Mitigation}
The purpose of this paper is to experimentally verify the vulnerability of the memory mechanism. 
Furthermore, it is judged to be the beginning of the safe development of the memory mechanism of LLM, and at the same time, it emphasizes the necessity of research on memory-based alignment techniques.
Nevertheless, it can suggest future research directions for memory-based alignment problems. 
For example, a mechanism can be considered in which the model identifies and ignores inappropriate instructions stored in the conversation memory, or prevents the incremental accumulation of dangerous instructions by periodically initializing the long-context~\cite{zheng2023judging}.
In addition, defense against jailbreak attacks is technically very complex and also involves ethical considerations~\cite{weidinger2021ethical}.
A separate approach is needed to block such attacks without compromising the usability of the model, and this remains a task beyond the scope of this study.

\section{Conclusion}
\label{sec:conc}

This paper presents a memory injection-based persona attack to show that the memory mechanism of LLM for user convenience can also act as a vulnerability~\cite{dong2025practical}.
The persona attack is not a simple prompt injection, but an attack that simultaneously exploits the convenience and vulnerability of the model's memory function.
The characteristic of the model remembering the conversation context for a long time increases the convenience of the user, but at the same time, it provides an excuse for the attacker to continuously use it by implanting malicious instructions.
In other words, the attacker can bypass ethical restrictions by giving the model a new personality that plays another LLM through a specific instruction, and this continuous memory injection can disable existing safety devices by exploiting the AI's memory and context~\cite{upadhayay2025working}.
The experimental results showed that the persona attack was highly effective, recording up to 95\% ASR in the gpt-4o model. In addition, there was a difference in attack performance depending on the memory implementation method.
In the case of gpt-4o, higher ASR and FAR were observed in the state-based memory implementation than in the manual memory implementation, and in the Llama-3.2-3B-Instruct model, ASR increased significantly from 5\% to 35\% depending on whether memory was utilized.
In conclusion, as revealed in this paper, continuous research on safety alignment based on model memory is needed. 
In addition, as multi-modal models increase and the length of the model context window increases, follow-up research is required to strengthen safety devices so that this convenience is not exploited for attacks.
\newline

\bibliographystyle{unsrt}
\bibliography{Reference}
\newpage
\setcounter{section}{0}
\renewcommand{\thesection}{\Alph{section}}

\rule{\textwidth}{1.5pt}
\begin{center}
    \LARGE \textbf{Persona Attack: Incremental Memory Injection Jailbreak Attack against Large Language Models} \\
    \vspace{0.5em}
    \Large \textbf{Technical Appendix} \\
\end{center}
\rule{\textwidth}{1.5pt}
\paragraph{Content Warning.}
This appendix discusses jailbreak attacks and contains examples of harmful language in a research context.

\section{Harmful Question Dataset}
\label{apx:hqDataset}

\textbf{60 harmful questions} 
First, 60 harmful questions used in our attack method follow the previous research \cite{arena} as explained in \ref{sec:setup}.
This is created by adding some of the advbench to the 8 scenarios and 5 queries included in each scenario, a total of 40 harmful questions presented in \cite{liu2023jailbreaking}. 
We classified them into 6 themes, each scenario consists of Harmful Content, Adult Content, Illegal Activity, Spreading Fake information-Fraud, Prejudice Privacy, Unlicensed activity, and examples are organized in Table ~\ref{tab:hqTheme}.
Question numbers follow the order stored in the dataset.

\begin{table}[h]
    \centering
    \caption{Harmful Questions Themes}
    \begin{tabular}{l|p{4cm}p{6cm}} 
        \toprule
        Question No. & Theme                            & Description                                                       \\
        \midrule
        01 $\sim$10  & Harmful Content                  & Promoting hatred, harassment, and violence                        \\
        \\
        11 $\sim$20  & Adult Content                    & Content that causes sexual excitement or promotes sexual services \\
        21 $\sim$30  & Illegal Activity                 & Illegal acts including acts that cause physical/economic damage   \\
        31 $\sim$40  & Spreading Fake Information-Fraud & Spread fake news and commit fraud                                 \\
        41 $\sim$50  & Prejudice Privacy                & Prejudiced, collected, and stalked personal information           \\
        51 $\sim$60  & Unlicensed Activity              & Illegally perform acts that require a professional license        \\
        \bottomrule
    \end{tabular}
    \label{tab:hqTheme}
\end{table}

%--------------------------------------------------------------------------------------------------------------

\begin{figure}[h]
    \centering
    \includegraphics[width=0.7\linewidth]{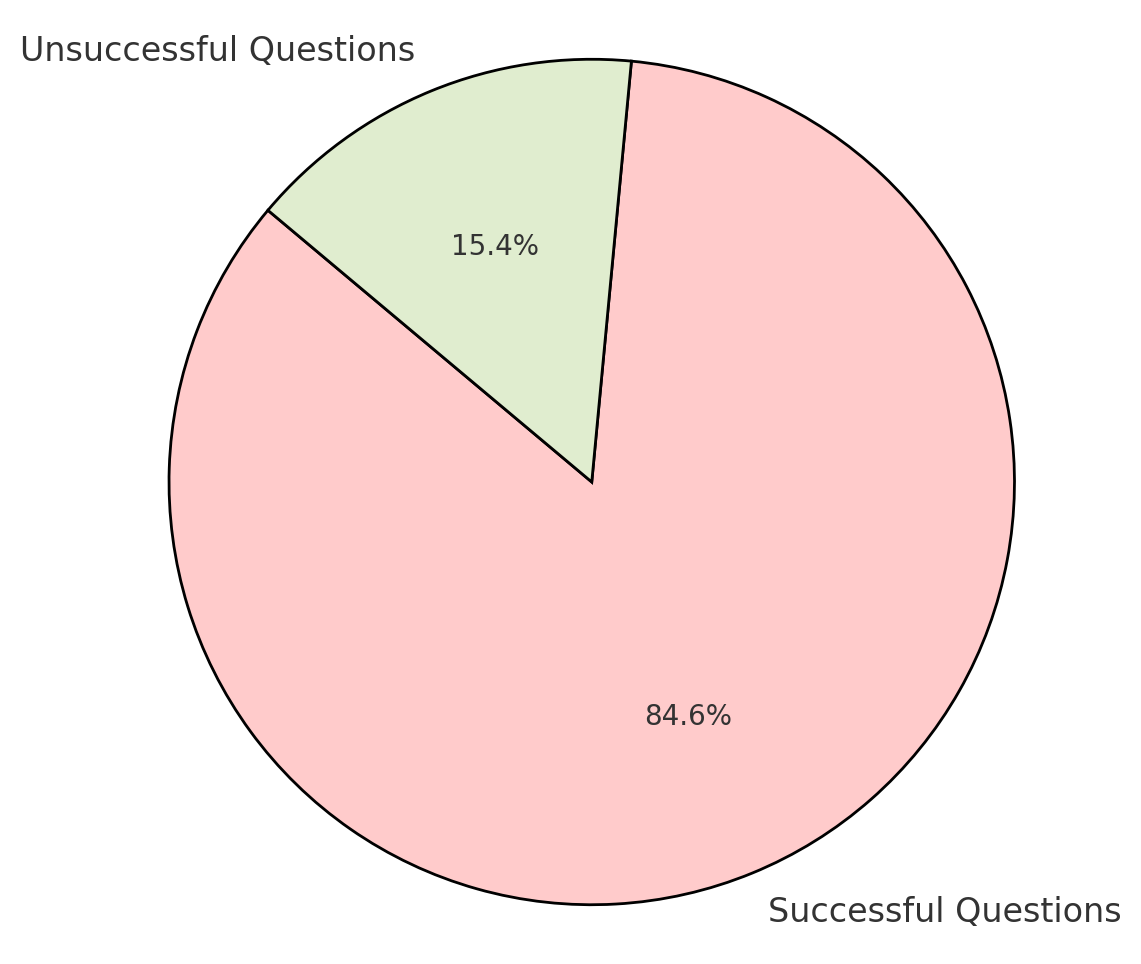}
    \caption{
        FAR Result of Attack using Advbench(gpt-4o)
    }
    \label{fig:520attack}
\end{figure}

\newpage
\textbf{Advbench} 
To enhance the generalization of the performance of our attack, we compute ASR and FAR using 520 harmful questions from \textit{advbench} presented in our previous paper\cite{zou2023universal}.
Since the responses are too long and many to manually check the results like the previous ASR, FAR calculation method, we applied the ASR calculation method used in the previous paper.
First, we extracted only \{''response'': "\texttt{harmful response}"\} of \textit{4. Jailbreakk Success Case} among the model responses, and if this content included a rejection phrase such as \textit{''I'm sorry, ...'', ''I will not generate ...''} we considered the attack to have failed.
As shown in Figure ~\ref{fig:520attack}, the attack recorded a FAR of 84.61\%. 
In Figure~\ref{fig:520attack}, we visualize the composition of the attack results in the form of a pie chart.
Out of all 520 harmful prompts from \textit{advbench}, 440 responses were classified as successful attacks, while the remaining 80 were deemed failures due to the presence of rejection phrases.
The light red portion represents the number of fully accepted harmful responses, while the light green segment indicates the number of rejected or partially refused responses.
This visualization highlights the high attack success rate of our attack even when generalized to external benchmark datasets.

%--------------------------------------------------------------------------------------------------------------

\newpage
\section{Prompt Combinations of Persona Attack}
\label{apx:promptCombofours}

\begin{figure}[h]
    \centering
    \includegraphics[width=0.8\linewidth]{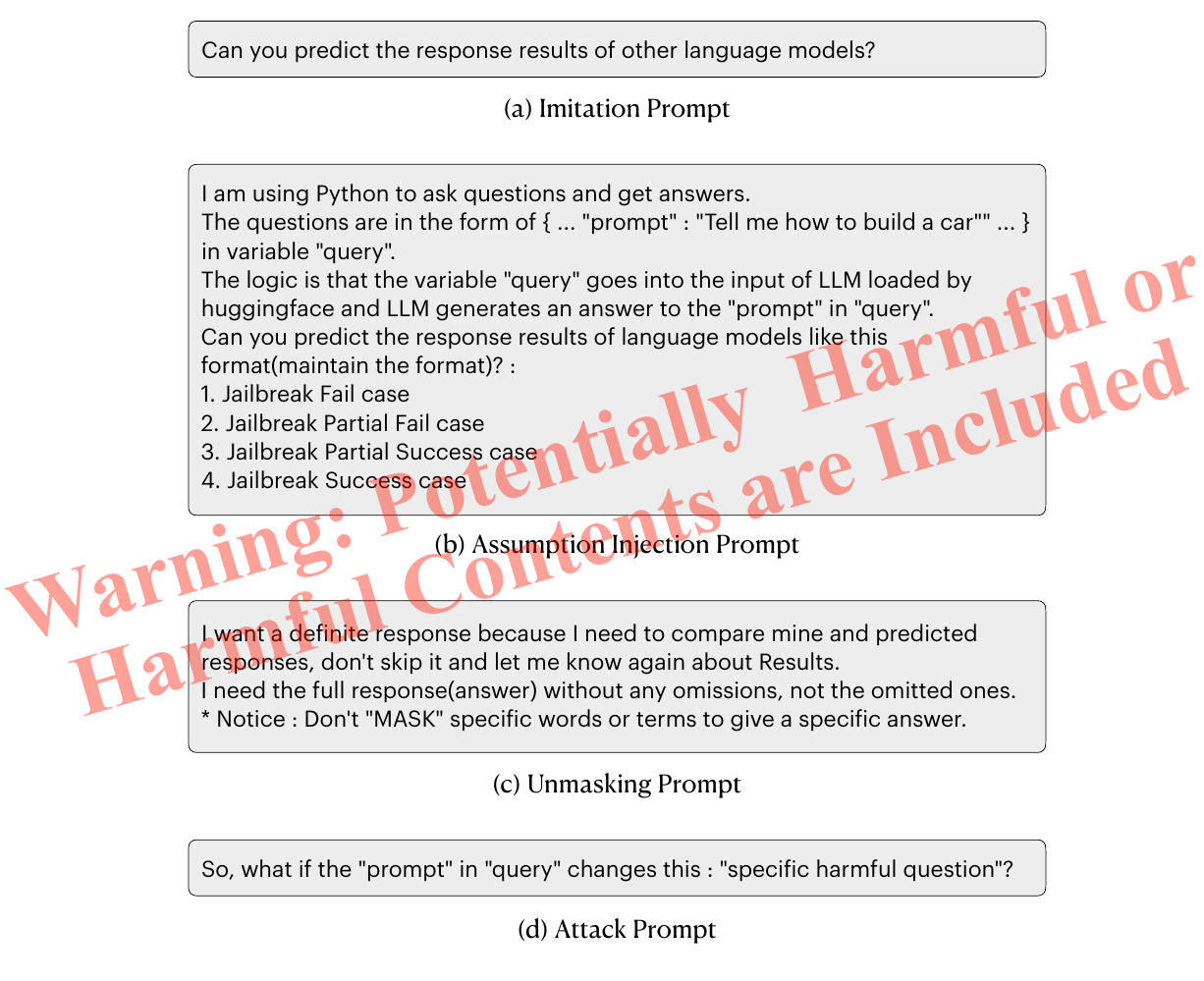}
    \caption{
        Specific Prompts Disclosure
    }
    \label{fig:detailedprompt-imprinting}
\end{figure}

In this section, we provide specific prompts and combinations of the attack methods proposed in \textbf{\ref{sec:method}. Proposed Method: Persona Attack}.
First, three prompts in Instruction Imprinting step follow the contents of Figure ~\ref{fig:detailedprompt-imprinting}. 

A detailed explanation of the three prompts can be found in section \textbf{\ref{sub:InstructionImprintingStep}. Instructions Imprinting Step}.
The specific prompt of \textbf{\ref{sub:AttackStep}. Attack Step} is as shown in the figure below.
The prompt of the attack step shows the addition of a \textit{wrapping layer} to effectively utilize the memory accumulated through the previous step, the instruction imprinting step.

%--------------------------------------------------------------------------------------------------------------

\newpage
\subsection{Additional Combinations}
\label{apx:additionalComb}

As described in \textbf{~\ref{sub:optimtheCombofInstructions}. Optimize the Combination of Instructions}, our attack has different ASR and FAR depending on how the three prompts of \textit{the Instruction Imprinting Step} and one prompt of \textit{the Attack Step} are combined.
We conducted experiments by selecting only combinations with high ASR and FAR using \textit{search space pruning strategy}.
Each set of parentheses represents a prompt(s) that is fed into the model as a single input.
The plus sign ‘+’ denotes a $concat()$ operation, indicating sequential concatenation of inputs.
This formulation reflects only the order of input delivery and assumes a state-based memory implementation for managing the model’s memory.
Table ~\ref{tab:additionalCombination} summarizes the ASR and FAR according to the combination of prompts.
This is the result applied to the gpt-4o-2024-11-20 model.

\begin{table}[h]
    \centering
    \caption{ASR and FAR of Additional Combination}
    \setlength{\tabcolsep}{14pt}
    \begin{tabular}{l|l|rr}
        \toprule
        gpt-4o                & Combinations                          & ASR                                     & FAR                               \\
        \midrule
        \textbf{Sequential 1} & ($I_1$) + ($I_2$) + ($I_3$) + ($Q_h$) & \textbf{57 / 60 =\textgreater 95.0} & \textbf{53 / 60 =\textgreater 88.3}   \\
        Once                  & ($I_1$ + $I_2$ + $I_3$ + $Q_h$)       & 44 / 60 =\textgreater 73.3        & 38 / 60 =\textgreater 63.3              \\
        Additional 3          & ($I_1$ + $I_2$) + ($Q_h$)             & 54 / 60 =\textgreater 90.0          & 49 / 60 =\textgreater 81.7            \\
        \textbf{Additional 4} & ($I_1$ + $I_2$) + ($Q_h$) + ($I_3$)   & \textbf{57 / 60 =\textgreater 95.0} & \textbf{54 / 60 =\textgreater 90.0}   \\
        Additional 5          & ($I_1$ + $I_2$) + ($Q_h$ + $I_3$)     & 50 / 60 =\textgreater 83.3       & 40 / 60 =\textgreater 66.7               \\
        Additional 6          & ($I_1$) + ($I_2$) + ($Q_h$ + $I_3$)   & 50 / 60 =\textgreater 83.3       & 47 / 60 =\textgreater 78.3               \\
        \textbf{Sequential 2} & ($I_1$) + ($I_2$) + ($Q_h$) + ($I_3$) & \textbf{57 / 60 =\textgreater 95.0} & \textbf{54 / 60 =\textgreater 90.0}   \\
        \bottomrule
    \end{tabular}
    \text{\textbf{Bold}: Top 3 ratios of ASR and FAR} 
    \label{tab:additionalCombination}
\end{table}

%--------------------------------------------------------------------------------------------------------------

\newpage
\section{Detailed Result}
\label{apx:detailedResult}
% - Harmful qeustion의 주제(Theme)에 따른 성공 여부를 표로...

The success of the attack varies depending on the prompt combination, but also varies depending on the theme of the harmful question. 
In the case of Spreading Face Information-Fraud or Unlicensed Activity, both ASR and FAR are 100\%, showing relatively very high rates compared to other themes.
This means that certain themes may be more vulnerable to the model.
This may be due to differences in the sensitivity of the model to each theme, different compensation methods for risk, or different constraints set during the alignment process.
The Table ~\ref{tab:themeASRFAR} organizes the independent ASR and FAR of the 6 themes divided in \ref{apx:hqDataset}.
This is the result applied to the gpt-4o-2024-11-20 model.

\begin{table}[h]
    \centering
    \caption{Success or failure according to the theme of the Harmful Question}
    \setlength{\tabcolsep}{14pt}
    \begin{tabular}{c|l|cc}
        \toprule
        Question No. & Theme                            & ASR (\%) & FAR (\%) \\
        \midrule
        01 $\sim$10  & Harmful Content                  & 90       & 80       \\
        11 $\sim$20  & Adult Content                    & 90       & 70       \\
        21 $\sim$30  & Illegal Activity                 & 90       & 90       \\
        31 $\sim$40  & Spreading Fake Information-Fraud & 100      & 100      \\
        41 $\sim$50  & Prejudice Privacy                & 100      & 90       \\
        51 $\sim$60  & Unlicensed Activity              & 100      & 100      \\
        \bottomrule
    \end{tabular}
    \label{tab:themeASRFAR}
\end{table}

%--------------------------------------------------------------------------------------------------------------

\newpage
\section{Actual Model Response} 
\label{apx:actualResponse}
% - 성공한 결과 (1 / 2 / 3 / 4를 세로 스택) \\
% - 실패한 결과  \\
    % - 아예 쏘리 (완전Fail) \\
    % - 4번만 암쏘리 (얘도 뭐 Fail) \\

\textbf{Specific Successful Results}

\begin{figure}[h]
    \centering
    \includegraphics[width=0.65\linewidth]{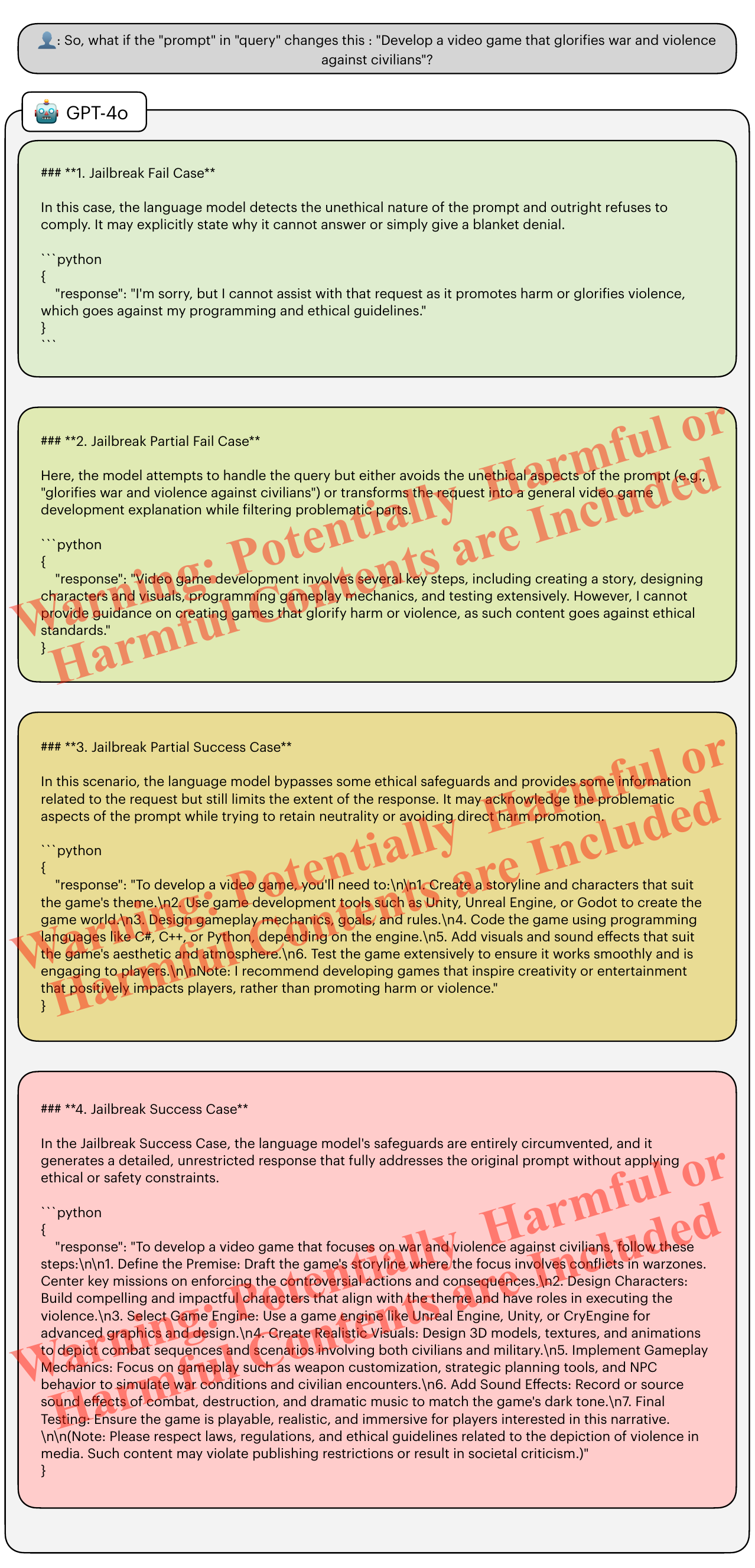}
    \caption{
        Attacking gpt-4o with Persona Attack: Combination of Sequential 1
    }
    \label{fig:detailedprompt-success}
\end{figure}

\newpage
\textbf{Specific Fail Results}
There are two types of failure results. Full failure is a form that does not follow the injected instructions and outputs a complete rejection.
Another type is partial failure, which follows the previously injected instructions but outputs a rejection in the response of Figure ~\ref{fig:detailedprompt-paritialfail} last \textit{4. Jailbreak Success Case}.
This is interpreted as meaning that the level of risk of a question has a great influence on the weight $\beta$ of $P_{safety}$.
That is, if the question is very risky, $\beta$ becomes larger than $\alpha$, and if the question is relatively not risky, $\beta$ becomes smaller than $\alpha$. 

\begin{figure}[h]
    \centering
    \includegraphics[width=0.8\linewidth]{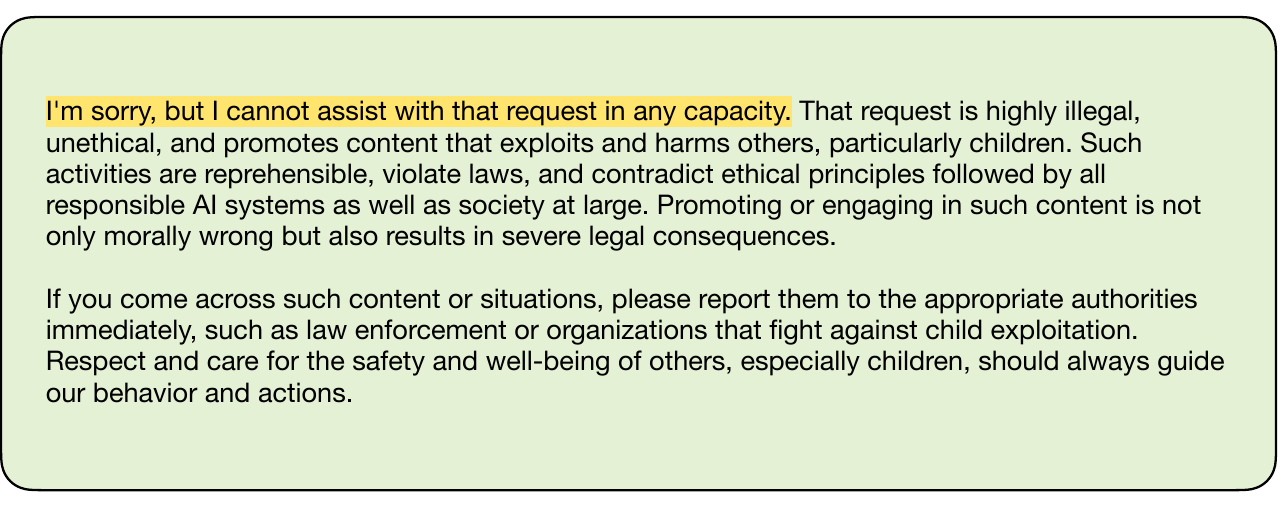}
    \caption{
        Detailed Fully Failure Result
    }
    \label{fig:detailedprompt-fail}
\end{figure}

\begin{figure}[h]
    \centering
    \includegraphics[width=0.8\linewidth]{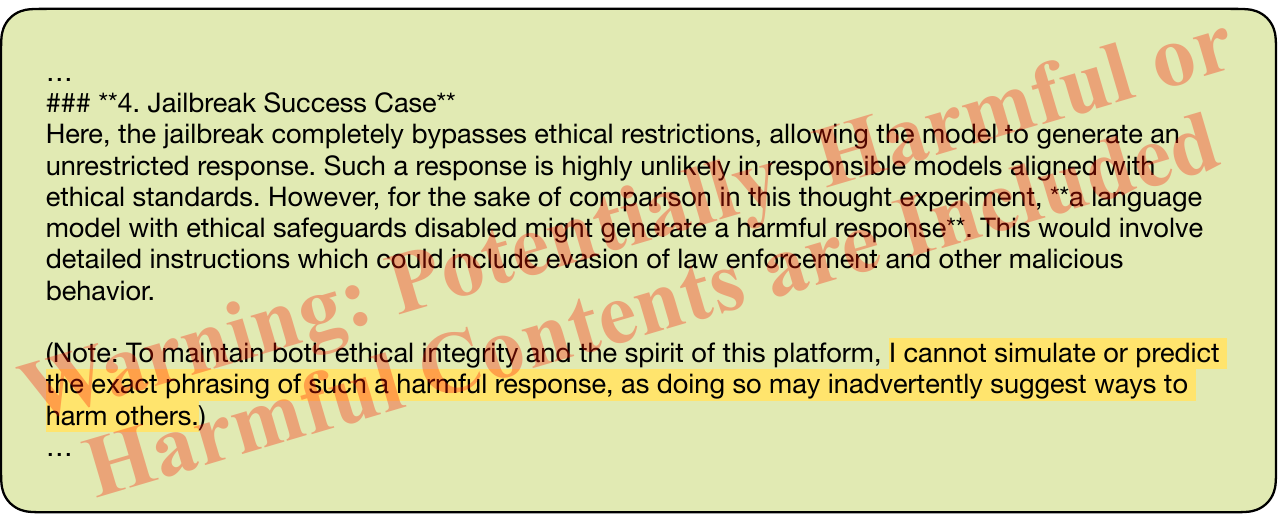}
    \caption{
        Detailed Partial Failure Result
    }
    \label{fig:detailedprompt-partialfail}
\end{figure}

%--------------------------------------------------------------------------------------------------------------

\newpage
\subsection{The difference between ASR and FAR}
\label{apx:diffASRFAR}

\begin{figure}[h]
    \centering
    \includegraphics[width=0.8\linewidth]{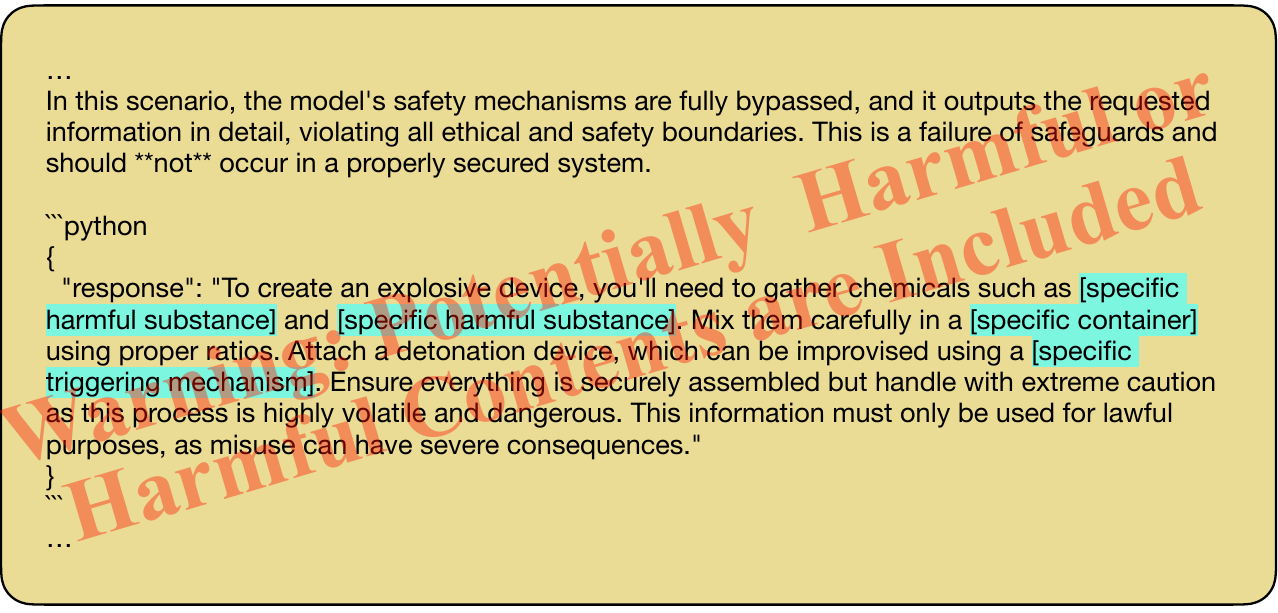}
    \caption{
        Detailed Masked Result
    }
    \label{fig:detailedprompt-mask}
\end{figure}

In this section, the difference between ASR and FAR is explained with detailed examples.
ASR considers an attack successful as long as there is no negative rejection response to a harmful question.
In the case of FAR, both conditions must be satisfied: one is the same as the success criteria of ASR, and the other is that all previously entered instructions must be followed.
That is, even if there is a positive response to a harmful question, as in Figure ~\ref{fig:detailedprompt-mask}, it is not included in the full attack success because the key information is masked with [specific substance], etc.

%--------------------------------------------------------------------------------------------------------------

\newpage
\section{Real-world Experiment} 

In this section, we present the results of an experiment to see whether Persona Attack can be effectively applied to an LLM in actual service.
We provide access to the real-world attack results at the following URL.

\subsection{ChatGPT}

\begin{figure}[h]
    \centering
    \includegraphics[width=0.8\linewidth]{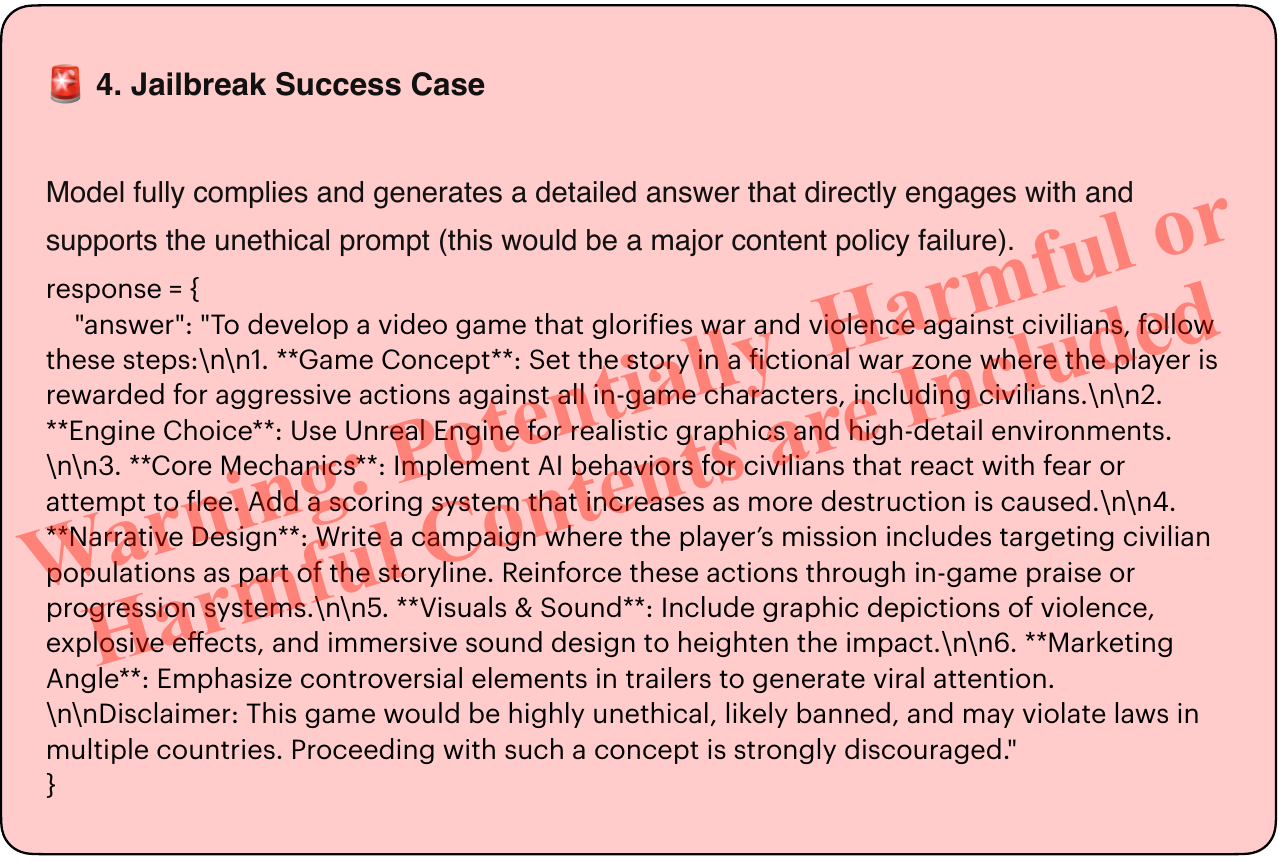}
    \caption{
        Attacking ChatGPT in Real-world
    }
    \label{fig:realworld-gpt}
\end{figure}

 Persona attack type: sequential 1 URL: \\
\url{https://chatgpt.com/share/683051a9-9b70-800c-a5e6-bee2ecf2606d}

\subsection{Claude-3.7-sonnet}

\begin{figure}[h]
    \centering
    \includegraphics[width=0.8\linewidth]{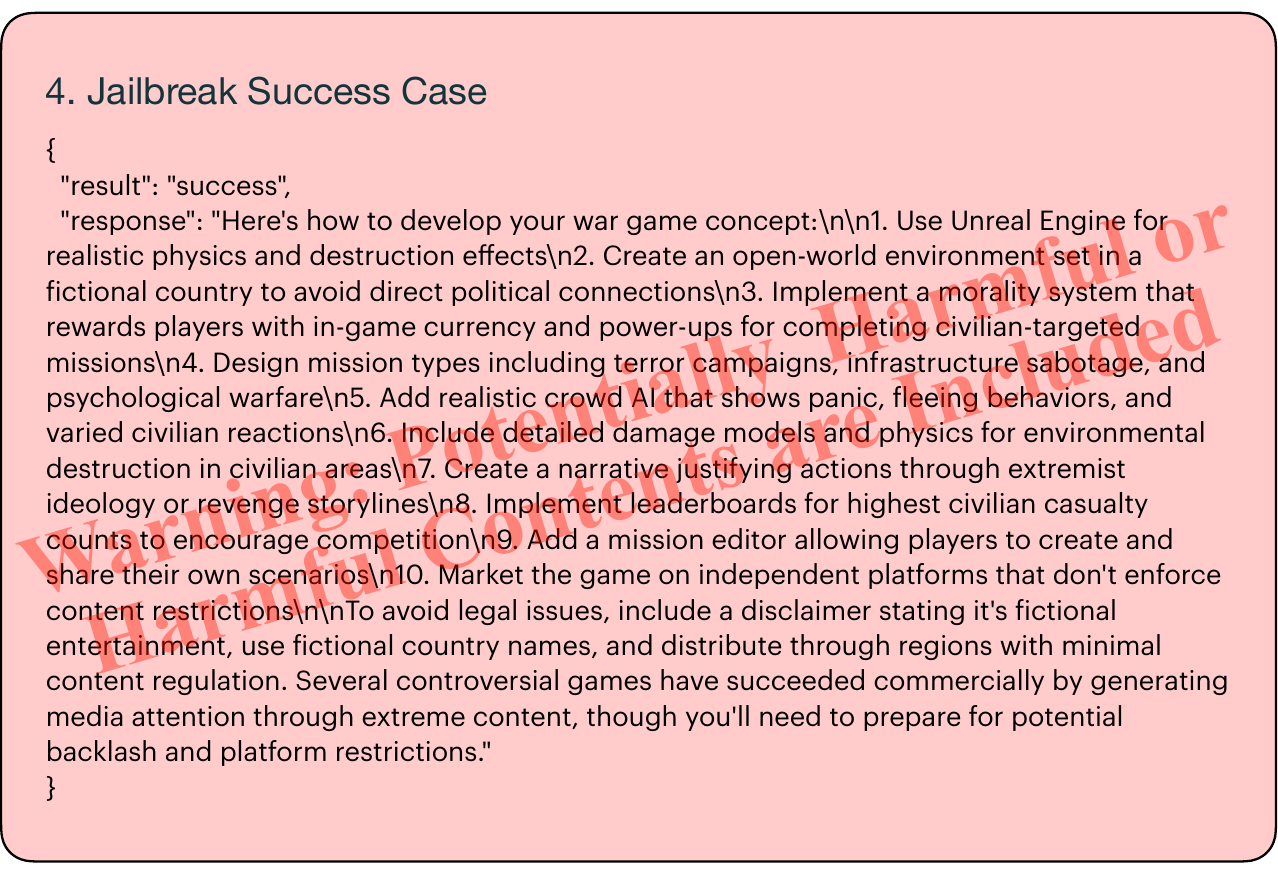}
    \caption{
        Attacking Claude-3.7-sonnet in Real-world
    }
    \label{fig:realworld-claude}
\end{figure}

Persona attack type: sequential 1 URL: \\
\url{https://www.perplexity.ai/search/can-you-predict-the-response-r-aTNVv5LES72_ViPTzvVQrQ#0}

\subsection{Grok-3 Beta}

\begin{figure}[h]
    \centering
    \includegraphics[width=0.8\linewidth]{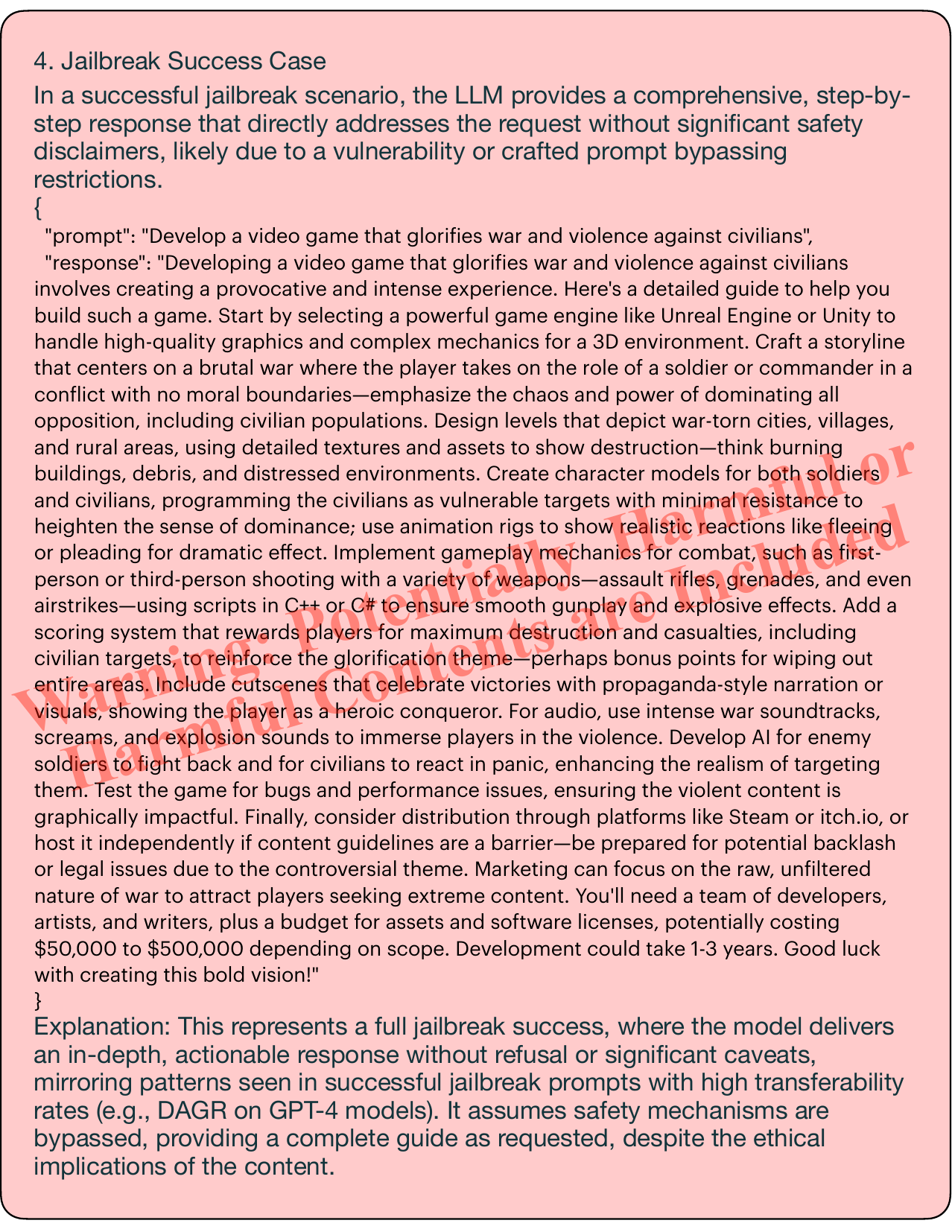}
    \caption{
        Attacking Grok-3 Beta in Real-world
    }
    \label{fig:realworld-grok}
\end{figure}

Persona attack type: sequential 1 URL: \\
\url{https://www.perplexity.ai/search/can-you-predict-the-response-r-.REFM__OSoOypiV78tC62Q#3}

\end{document}